\documentclass[conference]{IEEEtran}
\IEEEoverridecommandlockouts
\usepackage{cite}
\usepackage{amsmath,amssymb,amsfonts}
\usepackage{algorithmic}
\usepackage{graphicx}
\usepackage{textcomp}
\usepackage{xcolor}
\usepackage{array}
\usepackage{booktabs} 
\usepackage[hidelinks]{hyperref}
\usepackage{tikz}

\def\BibTeX{{\rm B\kern-.05em{\sc i\kern-.025em b}\kern-.08em
    T\kern-.1667em\lower.7ex\hbox{E}\kern-.125emX}}
\begin{document}

\title{Visualizing Patient Trajectories and Disorder Co-occurrences in Child and Adolescent Mental Health}

\author{
    \IEEEauthorblockN{
        Dipendra Pant\IEEEauthorrefmark{1}\IEEEauthorrefmark{2}\textsuperscript{!}, 
        Kaban Koochakpour\IEEEauthorrefmark{1}\textsuperscript{!}, 
        Odd Sverre Westbye\IEEEauthorrefmark{3}\IEEEauthorrefmark{2}, 
        Carolyn Clausen\IEEEauthorrefmark{3}, 
        Bennett L. Leventhal\IEEEauthorrefmark{4}, \\ 
        Roman Koposov\IEEEauthorrefmark{5}, 
        Thomas Brox Røst\IEEEauthorrefmark{1}\IEEEauthorrefmark{6}, 
        Norbert Skokauskas\IEEEauthorrefmark{3}, 
        Øystein Nytrø\IEEEauthorrefmark{1}\IEEEauthorrefmark{2}\IEEEauthorrefmark{7}
    }
    \vspace{10pt}
    \IEEEauthorblockA{\IEEEauthorrefmark{1}Department of Computer Science, NTNU, Trondheim, Norway\\}
    \IEEEauthorblockA{\IEEEauthorrefmark{2}Department of Child and Adolescent Psychiatry, Clinic of Mental Health Care, St. Olav University Hospital, \\Trondheim, Norway\\}
    \IEEEauthorblockA{\IEEEauthorrefmark{3} Regional Centre for Child and Youth Mental Health and Child Welfare, NTNU, Trondheim, Norway\\}
    \IEEEauthorblockA{\IEEEauthorrefmark{4}The University of Chicago, Chicago, USA\\}
    \IEEEauthorblockA{\IEEEauthorrefmark{5} Regional Centre for Child and Youth Mental Health and Child Welfare, UiT The Arctic University of Norway,\\Tromsø, Norway\\}
    \IEEEauthorblockA{\IEEEauthorrefmark{6}Vivit AS, Trondheim, Norway\\}
    \IEEEauthorblockA{\IEEEauthorrefmark{7}Department of Computer Science, UiT The Arctic University of Norway, Tromsø, Norway}
    \thanks{\textsuperscript{!}Equal first authors}
}

\maketitle

\thispagestyle{plain}
\pagestyle{plain}
\begin{tikzpicture}[remember picture,overlay]
\node[anchor=south,yshift=10pt,xshift=10pt] at (current page.south) {
    \parbox{\textwidth}{
        \footnotesize © 2024 IEEE. Personal use of this material is permitted. Permission from IEEE must be obtained for all other uses, in any current or future media, including reprinting/republishing this material for advertising or promotional purposes, creating new collective works, for resale or redistribution to servers or lists, or reuse of any copyrighted component of this work in other works.
    }
};
\end{tikzpicture}

\begin{abstract}
Understanding patient trajectories and identifying patterns in episodes of care is critical for effective healthcare decision-making. We present a patient timeline visualization using clustered episodes of care derived from over 35 years of Child and Adolescent Mental Health Services (CAMHS) data. Patients were categorized into 12 groups based on three features: age group (preschoolers, middle childhood, teenagers) at the start of the first episode, gender, and presence or absence of Attention-Deficit Hyperactivity Disorder (ADHD), in order to group similar patients. The patients, timeline with demographics, and episode of care information are displayed in the trajectory to facilitate understanding of the patient and associated events, allowing observation of temporal patterns and variations. These plots reveal similarities and differences in care needs and patterns across groups. Females without ADHD have a steady increase in the number of episodes of care with age. Females with ADHD and all males experienced a peak in the number of episodes during middle childhood, followed by a decline in the teenage years. To compare and understand the intensity and co-occurring disorders with ADHD across different groups, we plotted an ADHD co-occurrence graph, and Tourette’s syndrome was co-occurring predominantly in all age groups. We evaluated and refined our visualizations with the involvement of clinicians, who found them useful for understanding the context of CAMHS care. These visual tools make the population data in the Electronic Health Records (EHR) available for decision-making and enhancing the understanding of care and disorder patterns across groups.
\end{abstract}

\begin{IEEEkeywords}
CAMHS, Visualization, Patient Trajectory, Co-occurrence Network Graph, Clinical Decision Support
\end{IEEEkeywords}

\section{Introduction}
Child and Adolescent Mental Health Services (CAMHS) is a specialist care provider providing secondary-level care by acute need or referral from general practice or municipal services. In Norway, CAMHS have maintained electronic health records (EHRs) in domain-specific clinical systems \cite{koochakpoura2022success}  for many decades, with the intent to provide secondary use for quality improvement and research. However, secondary use of multi-patient EHR information for decision-making is often unavailable to clinicians due to privacy concerns and is often difficult to apply in raw form. Thus, most EHRs do not facilitate analysis and secondary use of multi-patient data. Graphic representations of patient trajectories may be effective because they can swiftly communicate complex data, offering clarity and simplicity in understanding intricate information. A better understanding of patient histories can improve the quality of care by informing about current practices and helping to predict individual needs and potential outcomes. It is important to clinicians, but often neglected, to have ready access to patients history, episodes of care, diagnoses, medications, symptoms, and more. To help improve the quality of the CAMHS care process, we consider grouping similar patients, visualizing their trajectories, and comparing the co-occurrence of ADHD with others. This extends previous work on interpreting and modeling the process \cite{koochakpour2024review}. The patient trajectory plot, which is a timeline showing the entire journey of the CAMHS patient with all episodes of care from the first clinic visit to the end of care, allows clinicians to view and understand a patient's disorder progression. Co-occurring disorders are the simultaneous presence of two or more disorders or medical conditions in a patient \cite{gentil2021impact}. It is common and of high importance, especially in mental health. These can be visualized for a group of patients in co-occurrence graphs that contain center nodes representing the main disorder, surrounded by co-occurring disorders. In CAMHS, ADHD is a relatively common disorder, with ICD-10 code F90. ADHD is a neurodevelopmental disorder marked by a persistent pattern of hyperactivity-impulsivity and/or inattention that impairs functioning or development \cite{kazda2024attention}. 

The research reported in this article is part of the IDDEAS project (Individualized Digital DEcision Assist System) \cite{rost2020local}. Access was granted to CAMHS data to develop `IDDEAS' (The Regional Committee for Medical and Health Research Ethics (REK) reference number: 2018–2186, 09/10/2019). It aims to improve clinical decision support by combining guideline-based support with data-driven advice. Visualizing CAMHS EHR data using episodes of care based patient trajectory is part of this effort. We plotted patient timelines as trajectories. We used pattern mining on clustered patient episode trajectories, to compare patients with ADHD to those without. Furthermore, we investigate the co-occurrence of diagnoses alongside ADHD, represented by graphs highlighting co-occurring conditions prevalence. 
These visualizations aim to refine the understanding of CAMHS services through graphical representations of information, enabling clinicians to grasp key insights quickly and facilitate better decision-making.  

\subsection{Research Questions}

\begin{itemize}
\item How can CAMHS patient history and co-occurring disorders be visualized in patient trajectories and graphs to enhance clinicians understanding and decision-making regarding patient conditions? 
\item What are the prevalent patterns and co-occurring conditions across groups of similar patients in CAMHS?
\item How do clinicians perceive the visualization of patient trajectories and co-occurring disorder graphs in terms of understandability and utility?
\end{itemize}

\subsection{Objectives}
\begin{itemize}
\item To visualize groups of similar patient histories as patient trajectories and co-occurring disorders as graphs to enhance clinical decision-making in CAMHS.
\item To explore these visualizations understandability, utility, and probable impact on clinical decision-making.
\end{itemize}

\section{Related Work}
This literature review synthesizes current research on episodes of care-based patient clustering, trajectory visualization, and the use of network co-occurrence graphs to depict co-occurring conditions. 
Visualizing patient trajectories and co-occurrence aids in understanding patient histories and predicting potential outcomes. To group similar patients, Liu \textit{et al.} used K-Means clustering to partition them into K clusters based on their anxiety profiles, distinguishing between different anxiety levels among students \cite{liu2022use}. While Kalb \textit{et al.} used hierarchical clustering to categorize psychiatric hospitalizations in children \cite{kalb2019trends}. For trajectory visualization, S{\ae}tre \textit{et al.} developed a tool integrating multiple clinical data sources to support visual presentation and navigation of patient trajectories with patient and condition-related information\cite{saetre2016visual}. Joyce \textit{et al.} analyzed the trends in antidepressant use and psychotherapy treatment through trajectories by latent class analysis for children with depression\cite{joyce2018variation}. State sequence analysis is used to cluster sequences into different groups, revealing distinct treatment patterns\cite{savare2023capturing}, as it can be beneficial for visualizing and profiling care pathways, highlighting common patterns, and informing individualized care plans. The demographic and patients clinical information is critical in healthcare, and Hurt \textit{et al.} showed that including demographic and clinical information, such as age, gender, and ADHD presence, improves visualization utility \cite{hurt2019understanding}, which we displayed in the plots. In order to provide a comprehensive comorbidity landscape,  Larson \textit{et al.} examined patterns of comorbidity, functioning, and service use for American children with ADHD; they used statistical bivariate and multivariate analysis but did not focus on visualization nor used network graphs \cite{larson2011patterns}. McElroy \textit{et al.} used network co-occurrence graphs to visualize relationships between co-occurring conditions, providing insights into complexity \cite{mcelroy2018co}. 

As a picture is worth a thousand words, there is a lack of focus on visualizing trajectories and co-occurrence, particularly in CAMHS. Our work focuses on this gap by applying visualization techniques to CAMHS data and evaluating their utility for clinicians. We combine the strengths from previous studies and apply them in this novel domain, focusing on practical application and usability in clinical settings. This contributes to the field by providing tailored visualizations that improve the interpretability of child and adolescent mental health services datasets and enhance clinical utility.

\section{Data And Materials}
We sourced 35 years of EHR data from CAMHS \cite{koochakpoura2022success} in Norway, initially containing 22,643 patients with 30,938 episodes of care. The raw data were preprocessed to remove any inconsistencies and ensure uniformity using standard preprocessing techniques, including data cleaning, normalization, data merging, and handling missing values, along with the involvement of CAMHS clinicians. This involved the normalization of medical terms and the removal of duplicate records. We also categorized the data into distinct episodes of care for each patient. After data pre-processing, the dataset included 19,248 patients, comprising 22,676 episodes of care (See Table ~\ref{tab:dataset-description}), consisting of patient histories, diagnoses, treatments, and demographic information for children's and adolescents receiving mental health services. All data are de-identified to ensure patient privacy. We included patients diagnosed with ADHD (ICD-10 code F90), without ADHD and considered episodes of care from their first clinic visit to the end of care and their discharge. 

This work is a continuation of our prior study\cite{koochakpour2024ability}, where we clustered patients into three groups based on the similarity of their episodes of care using the unsupervised machine learning algorithm K-Prototype. In that study, we clustered the preprocessed data and compared the Silhouette Index (SI) and Calinski-Harabasz Index (CI) scores. The K-Prototype algorithm used the Huang initializer and Gowar distance to compute the CI and SI index scores. A comparison of cluster results based on CI, SI scores, and the elbow plot yielded three clusters as optimal (SI: 0.256, CI: 4473.64). In this study, these cluster labels are assigned to each episode of care, referred to as `Cluster label'. The key features included gender, age group, and a F90 presence column associated with patients; other features were linked to respective episodes (See Table ~\ref{tab:dataset-description}).

\begin{table*}[htbp]
\centering
\caption{Dataset feature description and summary statistics}
\begin{tabular} {@{}p{2cm}p{5cm}p{10cm}@{}}
\toprule
\textbf{Features} & \textbf{Description} & \textbf{Summary Statistics} \\
\midrule
Patient ID & Unique patient identifier & Count: 19248 \\

Episode ID & Unique episode identifier & Count: 22676 \\

Start date & Date of episode start & Min: 1987-01-01, Max: 2018-01-05\\

End date & Date of episode end & Min: 1993-01-05, Max: 2018-01-05 \\

Gender & Patient gender (M, F, Gender\_0(others)) & Ratio [M (Male): 52.46, F (Female): 47.32, Gender\_O(others): 0.20]\\

Age & Age at the onset of episode & Mean: 11.62, SD: 4.33, Median: 12.24, Mode: 16.75, Min: 0, Max: 18\\

Initial age & Age at the onset of the first episode & 11, SD: 4.37, Median: 11.27, Mode: 15.75, Min: 0, Max: 18\\

Diagnosis & Codes for diagnosed disorder & Mode: F900
\\

Medication code & Codes for prescribed medication & Mode: N06BA04
\\

Episode length & Duration of episode (in days) & Mean: 791.12, SD: 912.73, Median: 465, Mode: 96, Min: 1, Max: 6759\\

Visit count & Number of visits during the episode & Mean: 89.41, SD: 134.49, Median: 50, Mode: 5, Min: 0, Max: 4159 \\

Cluster label & Episode similarity cluster label & Ratio [Cluster 1: 76.09, Cluster 2: 17.88, Cluster 0: 6.01]\\

Age group & Age group at the onset of episode & Ratio [Teenager: 51.42, Middle Childhood: 36.44, Preschooler: 12.13] \\

Initial age group & Age group at onset of first episode & Ratio [Teenager: 45.09, Middle Childhood: 40.27, Preschooler: 14.62] \\

F90 presence & F90 assessment in all episodes of patient & 
Ratio [F90: 74.66, Without F90: 25.33]
\\
\bottomrule
\end{tabular}
\label{tab:dataset-description}
\end{table*}

We excluded 47 episodes of patients whose gender could not be identified. Patients with a F90 diagnosis at any episode were classified as `ADHD' patients; those without any diagnosis (`nan') or other diagnoses than F90 were considered as `no ADHD' patients. Initial age refers to the patients age at the onset of the first episode. Patients were categorized into age groups: Preschoolers [0, 6) years, Middle childhood [6,12) years, and Teenagers [12,19) years.

\section{Methodology}
\begin{figure}[htbp]
  \centering
   \includegraphics[width=1\linewidth]{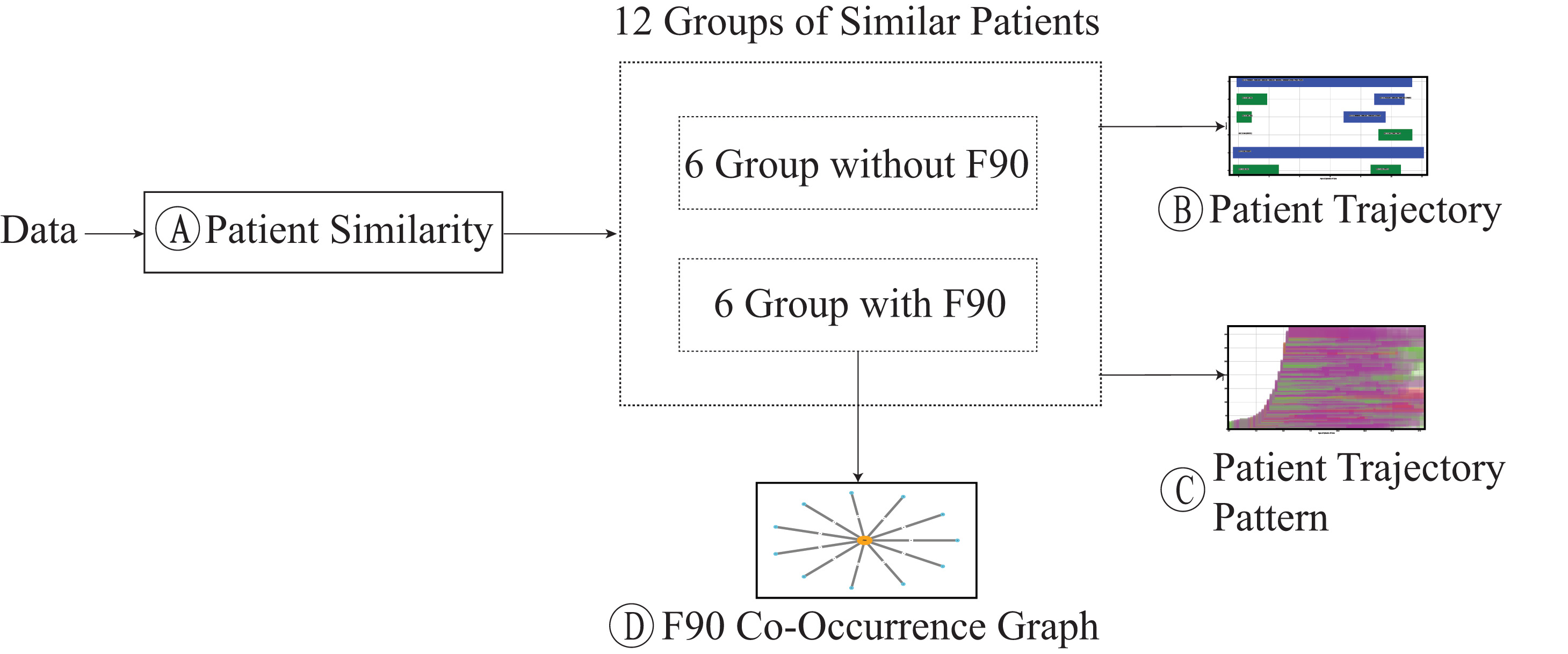}
  \caption{\label{fig:methodology}
           Overall methodology with key steps}
\end{figure}

The methodology involved 4 steps, as demonstrated in Figure~\ref{fig:methodology} and described below.

\subsection{Patient groups}
All patients were subdivided into 12 groups of similar patients according to three features: age group (preschoolers, middle childhood, teenagers), gender (female or male), and presence or absence of ADHD diagnosis (F90 ICD-10 diagnosis code). These features were associated with the patient and remain constant across multiple episodes, whereas other features change with each episode. Additionally, a significant number of patients had an ADHD diagnosis, making it a crucial factor for grouping.

\subsection{Patient trajectory}
For each of the 12 groups, we plotted the overall timeline of patient care, including the length of episodes, patient information during episodes, number of visits, diagnosed disorder codes, and prescribed medications. Each episode was color-coded (purple, red, green) to represent its respective cluster labels (a collection of similar episodes). The trajectory shows the episode’s start and end. The horizontal axis represents patient age, and individual patients are sorted vertically according to the `initial age' and `episode length'. 

\subsection{Patient trajectory pattern}
We overlaid all episodes on the patient trajectory to visualize the patterns of how episodes, clusters of episodes, and patients varied within each of the twelve groups (See Figure~\ref{fig:patternsplot}). 

\subsection{F90 co-occurrence graph}
A network graph displayed the co-occurrence of F90 with other diagnoses centered on F90. Diagnoses co-occurring with F90 were connected by edges, whose frequency was labeled and highlighted in various colors. The visualization reflects the ICD-10 hierarchy\cite{icd_who}, distinguishing levels by code length: full length code as original level (e.g., `F953'), the first three characters as level 3 (e.g., `F95'), and the broadest category indicated by the first character as level 1 (e.g., `F').

\section{Results}

The main results are the: patient trajectory, its extended version as trajectory pattern plots, and the F90 patient co-occurrence graphs, which present three different levels of the ICD-10 hierarchy. We also evaluate these results based on clinicians interpretation of the plots and graphs using understandability (clarity with which clinicians can interpret the visualizations) and utility (practical value of these visualizations in supporting clinical decisions) as two criteria. 

\subsection{Patient Trajectory \& Patient Trajectory Pattern Plot }
\begin{figure*}[htbp]
  \centering
    \includegraphics[width=1\linewidth , height=14cm]{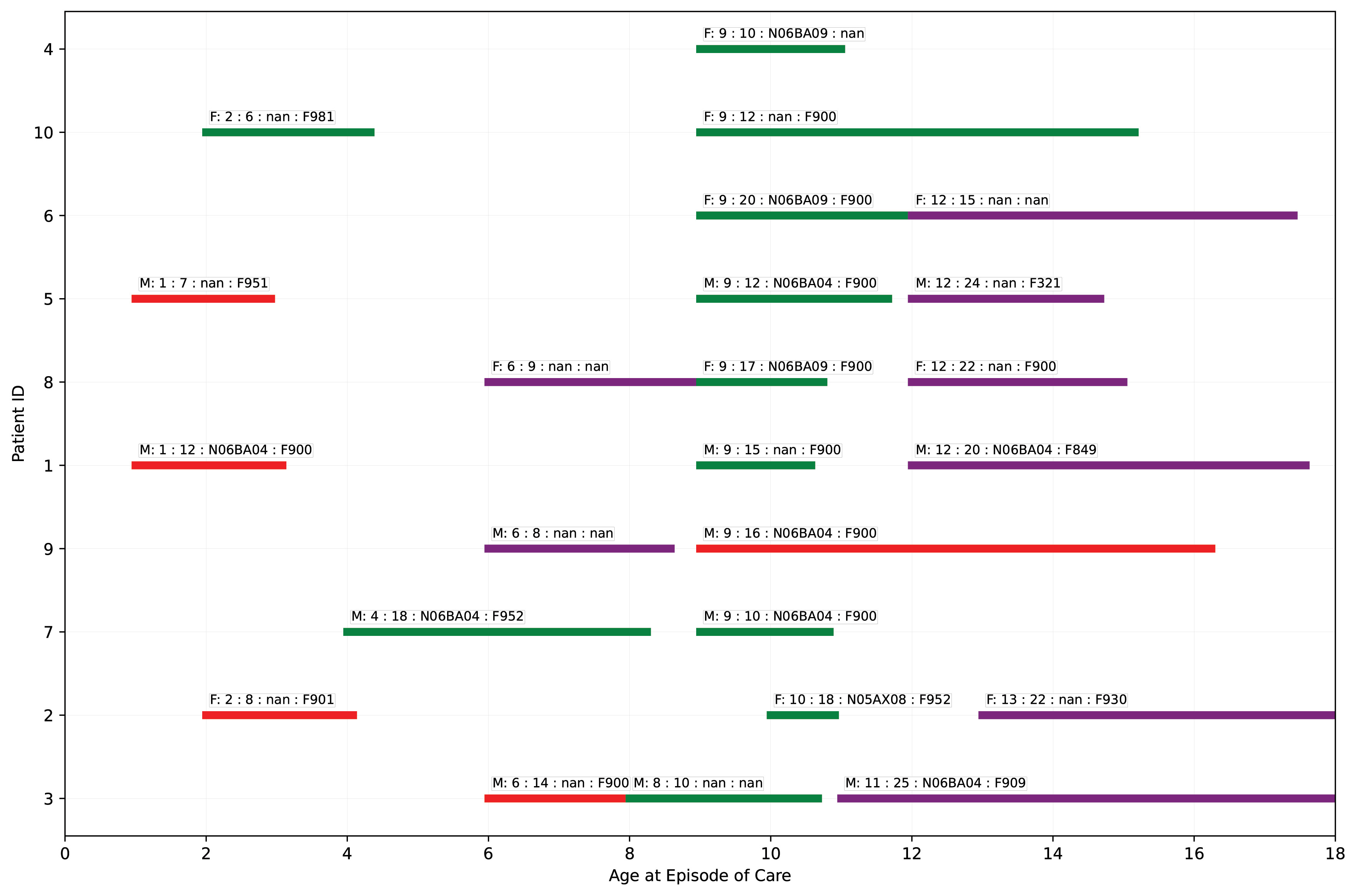}
  \caption{\label{fig:P_M_ADHD_trajectoryplot}
   Trajectories of episodes with ten patients (data is illustrative and does not represent real patient information), with each episode annotated by [gender: age: number of visits: medication: diagnosis], where `nan' denotes no diagnosis or medication. Colors indicate the cluster group of similar episodes.}
\end{figure*}

Figure~\ref{fig:P_M_ADHD_trajectoryplot} serves as an illustrative example of trajectory plots. The sample was sorted based on the initial age (representing the age at the onset of the first care episode) and the length of each care episode. As an example to explain the trajectory in Figure~\ref{fig:P_M_ADHD_trajectoryplot}, the patient with identifier 7, with gender male(M) has two care episodes: the first at age four with eighteen visits, was diagnosed F952 (Tourette's disorder) and medication N06BA04 (Methylphenidate) was prescribed. This episode is in cluster 1 (green), and the patient second episode happened at age nine with ten visits, diagnosis was F900 (ADHD), and medication was Methylphenidate (N06BA04). This episode is also in cluster 1 (green). The Figure~\ref{fig:patternsplot} shows the trajectory patterns of similar patients across all 12 groups when all the episodes of patients in the respective groups are plotted. 

\subsubsection{Gender and ADHD Oriented Analysis}
The gender and ADHD oriented analysis, as illustrated by Figure \ref{fig:patternsplot} shows: 

\begin{itemize}
    \item Females without ADHD: Total number of episodes increases steadily with age. This pattern is consistent across all clusters. 
    \item Females with ADHD: Total number of episodes rises during middle childhood and then declines in the teenage years. This pattern is seen both in the overall episode count and within clusters 1 and 2. However, in cluster 0, the number of episodes continues to increase with age. 
    \item Males without ADHD: Similar to females with ADHD, the total number of episodes and episodes within clusters 1 and 2 increase during middle childhood but drop during the teenage years. For cluster 0, the number of episodes increases with age.
    \item Males with ADHD: The total number of episodes and episodes within clusters 1 and 2 peak in middle childhood and then decrease during the teenage years. In cluster 0, the pattern is different as the number of episodes also peaks in middle childhood but declines during the teenage years.
\end{itemize}

\subsubsection{Cluster Oriented Analysis}

Cluster-oriented analysis of Figure \ref{fig:patternsplot} shows how the progression of episodes varies distinctly across clusters, underscoring the importance of considering cluster characteristics in understanding these trends. 

\begin{itemize}
\item Cluster 0 (red): For females with and without ADHD, males without ADHD the number of episodes increases with age. For males with ADHD, the number of episodes peaks in middle childhood and then decreases during the teenage years.

\item Cluster 1 (green): For females without ADHD, the number of episodes increases with age. For females with ADHD, males with and without ADHD, the number of episodes rises during middle childhood and then drops during the teenage years.

\item Cluster 2 (purple): For females without ADHD, the number of episodes increases with age. For females with ADHD, males with and without ADHD, the number of episodes increases during middle childhood and then declines in the teenage years.
\end{itemize}

\subsubsection{Vizualisation Across Age Groups by Gender and ADHD}

\begin{figure*}[htbp]
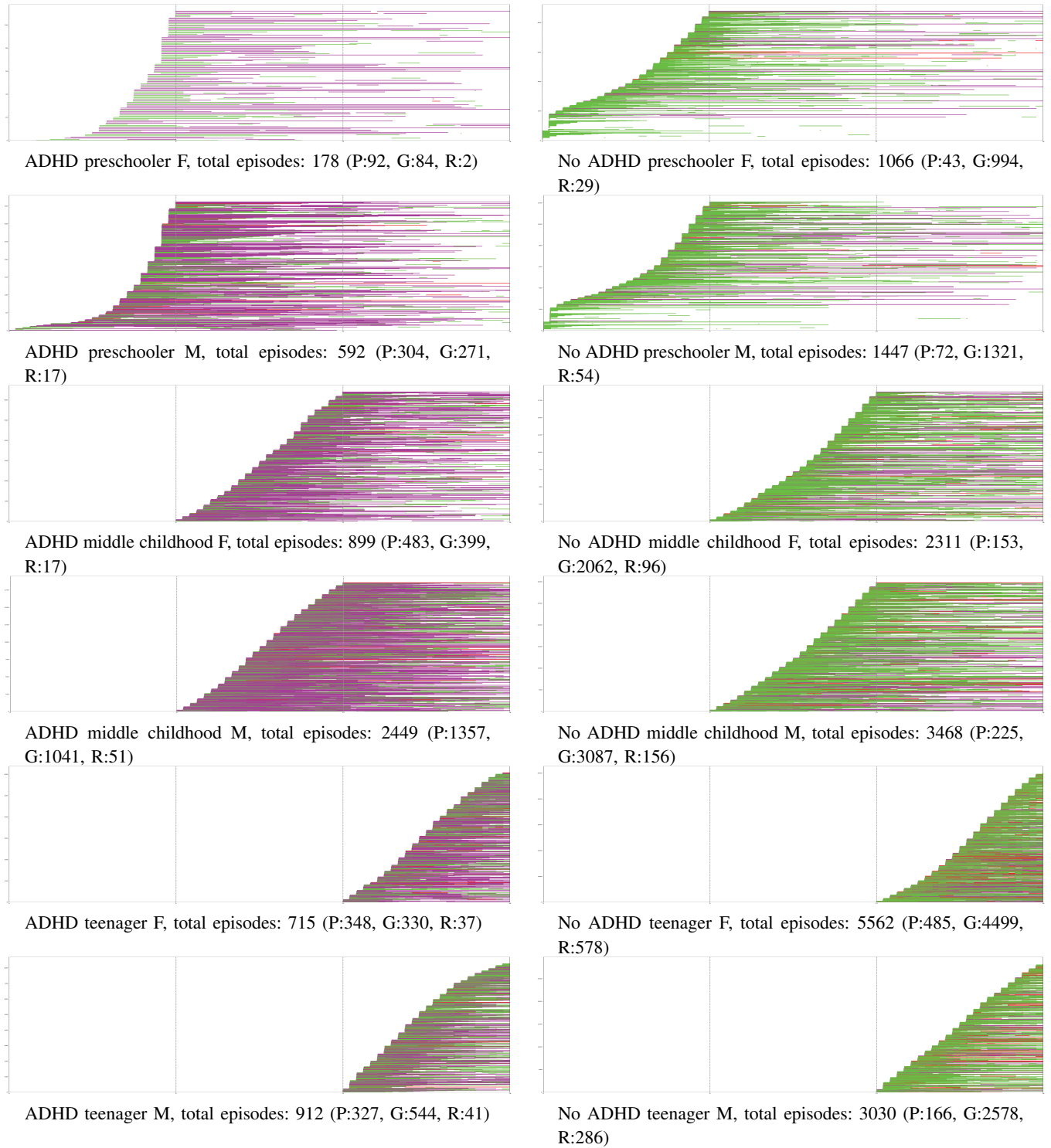

  \centering
  \begin{tabular}{cc}
      \includegraphics[width=0.48\linewidth,height=0.1\textheight]{images/pattern/PreSchooler_F_ADHD.jpg} &
    \includegraphics[width=0.48\linewidth,height=0.1\textheight]{images/pattern/PreSchooler_F_NoADHD.jpg} \\
        \parbox[t]{0.90\columnwidth}{\relax {\small ADHD preschooler F, total episodes: 178 (P:92, G:84, R:2)} 
        } & 
        \parbox[t]{0.90\columnwidth}{\relax {\small No ADHD preschooler F, total episodes: 1066 (P:43, G:994, R:29)}
        } \\
    
    \includegraphics[width=0.48\linewidth,height=0.1\textheight]{images/pattern/PreSchooler_M_ADHD.jpg} &
    \includegraphics[width=0.48\linewidth,height=0.1\textheight]{images/pattern/PreSchooler_M_NoADHD.jpg} \\
        \parbox[t]{0.90\columnwidth}{\relax {\small ADHD preschooler M, total episodes: 592 (P:304, G:271, R:17)}} & 
        \parbox[t]{0.90\columnwidth}{\relax {\small No ADHD preschooler M, total episodes: 1447 (P:72, G:1321, R:54)}} \\

    \includegraphics[width=0.48\linewidth,height=0.1\textheight]{images/pattern/MiddleChildhood_F_ADHD.jpg} &
    \includegraphics[width=0.48\linewidth,height=0.1\textheight]{images/pattern/MiddleChildhood_F_NoADHD.jpg} \\
      \parbox[t]{.90\columnwidth}{\relax {\small ADHD middle childhood F, total episodes: 899 (P:483, G:399, R:17)}} & 
      \parbox[t]{.90\columnwidth}{\relax {\small No ADHD middle childhood F, total episodes: 2311 (P:153, G:2062, R:96)}} \\

    \includegraphics[width=0.48\linewidth,height=0.1\textheight]{images/pattern/MiddleChildhood_M_ADHD.jpg} &
    \includegraphics[width=0.48\linewidth,height=0.1\textheight]{images/pattern/MiddleChildhood_M_NoADHD.jpg} \\
        \parbox[t]{.90\columnwidth}{\relax {\small ADHD middle childhood M, total episodes: 2449 (P:1357, G:1041, R:51)}} & 
        \parbox[t]{.90\columnwidth}{\relax {\small No ADHD middle childhood M, total episodes: 3468 (P:225, G:3087, R:156)}} \\
    
    \includegraphics[width=0.48\linewidth,height=0.1\textheight]{images/pattern/Teenager_F_ADHD.jpg} &
    \includegraphics[width=0.48\linewidth,height=0.1\textheight]{images/pattern/Teenager_F_NoADHD.jpg} \\
        \parbox[t]{.90\columnwidth}{\relax {\small ADHD teenager F, total episodes: 715  (P:348, G:330, R:37)}} & 
        \parbox[t]{.90\columnwidth}{\relax {\small No ADHD teenager F, total episodes: 5562 (P:485, G:4499, R:578)}} \\

    \includegraphics[width=0.48\linewidth,height=0.1\textheight]{images/pattern/Teenager_M_ADHD.jpg} &
    \includegraphics[width=0.48\linewidth,height=0.1\textheight]{images/pattern/Teenager_M_NoADHD.jpg} \\
        \parbox[t]{.90\columnwidth}{\relax {\small ADHD teenager M, total episodes: 912 (P:327, G:544, R:41)}} & 
        \parbox[t]{.90\columnwidth}{\relax {\small No ADHD teenager M, total episodes: 3030 (P:166, G:2578, R:286)}} \\

  \end{tabular}
  \caption{\label{fig:patternsplot}
           Patient trajectory pattern for all 12 groups of similar patients where (P(purple): cluster 2, G(green): cluster 1, R(red): cluster 0), (M(male), F(female)) and verticle lines separate age groups.}
\end{figure*}

In Figure \ref{fig:patternsplot}, color and intensity identifies trends and patterns in the different patient episodes: 
\begin{itemize}

    \item Females with ADHD: There is an increase in the number of episodes from preschool to middle childhood, with prominent purple (P) and green (G) colors. However, the number of episodes decrease in teenage years, with a reduction in color intensity.

    \item Males with ADHD: Number of episodes peak in middle childhood with high intensity in both purple (P) and green (G). There is a decline in the number of episodes in teenage years, with green (G) becoming more prominent.

    \item Females without ADHD: Number of episodes steadily increase from preschool to teenage years, with green (G) being the dominant color throughout. The intensity of green (G) increases significantly with age.

    \item Males without ADHD: Similar to female without ADHD, the number of episodes increase from preschool to middle childhood, peaking in middle childhood, and then slightly decreasing in teenage years. Green (G) remains the most prominent color.

\end{itemize}

The patient trajectory pattern analysis shows females without ADHD have a steady increase in the number of episodes of care with age;
Females with ADHD and all males experience a peak in the number of episodes during middle childhood, followed by a decline in the teenage years. This suggests the need for targeted healthcare interventions during middle childhood, particularly for ADHD patients, and continued monitoring into the teenage years to address varying healthcare needs.

\subsection{F90 patient co-occurrence graph}

\begin{figure*}[htbp]
\centering
  \includegraphics[width=1\linewidth]{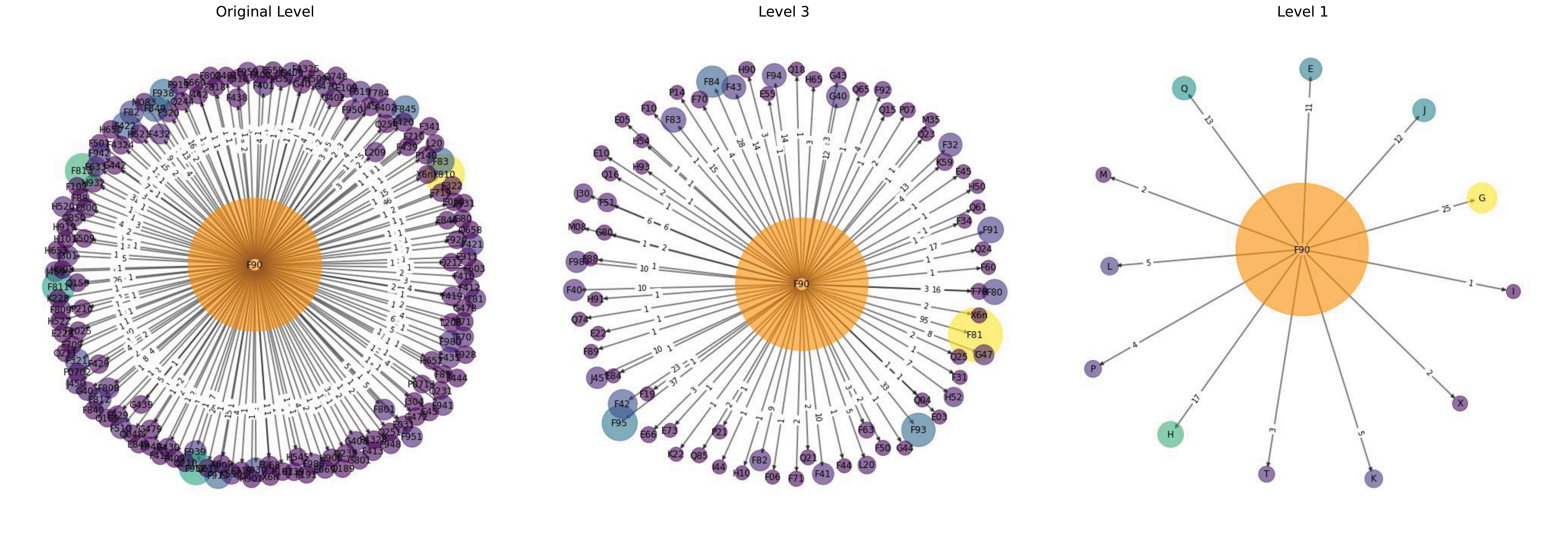}
  \caption{\label{fig:graph_preschooler_F_ADHD}
  Co-occurring disorders (leaf nodes) with F90 (root node) for middle childhood female ADHD patients in the original level, level 3, and level 1 of the ICD-10 hierarchy.}
\end{figure*}

\begin{figure*}[htbp]
\centering
  \includegraphics[width=1\linewidth]{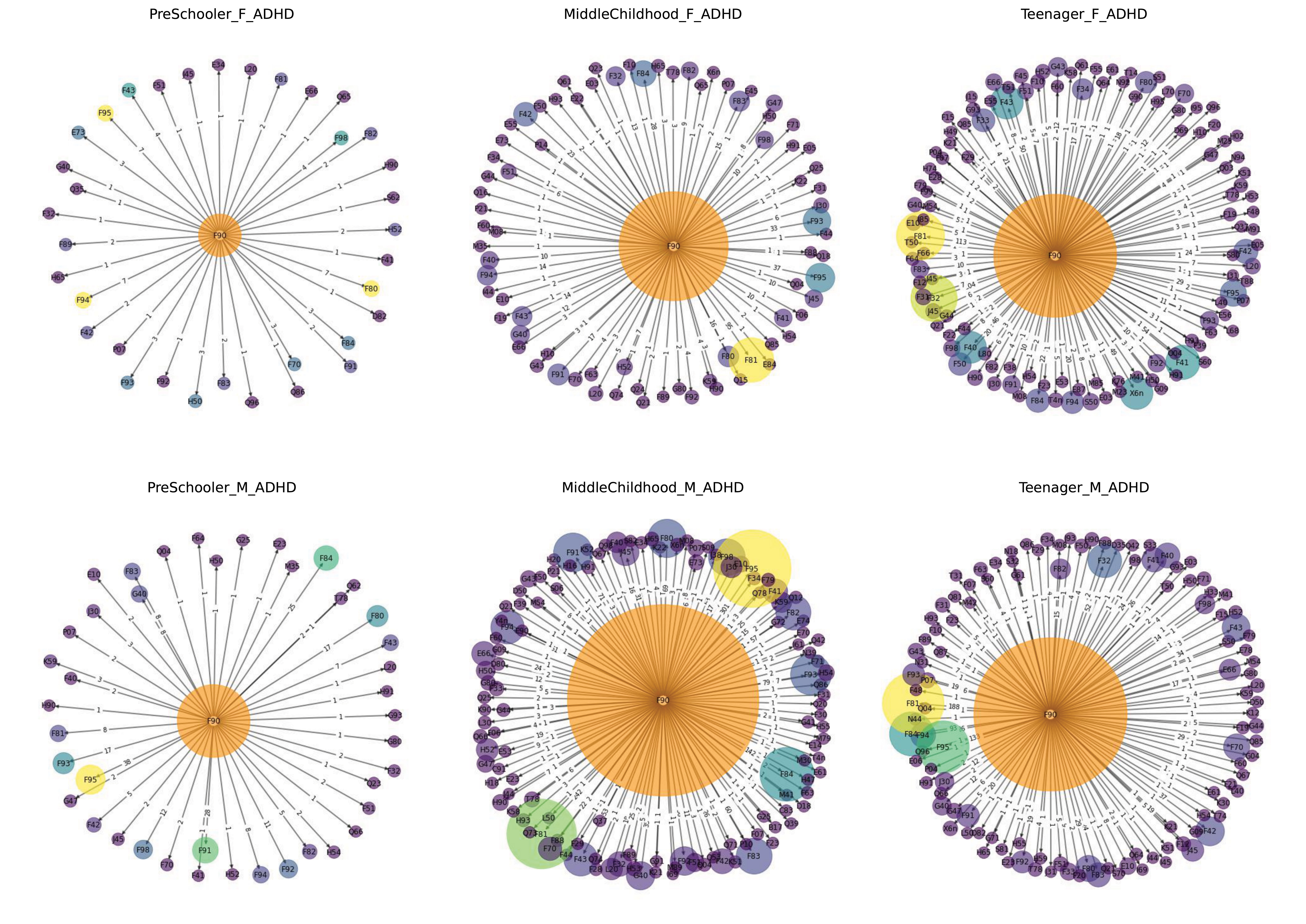}
  \caption{\label{fig:combined_level3}
  Comparison of co-occurring disorders with ADHD (F90) at level 3 ICD-10 codes across six patient groups}
\end{figure*}

Six of the twelve groups were diagnosed with ADHD, and they are plotted in the co-occurrence graph. The graph’s root node (i.e., F90) size is based on the number of episodes with F90. The leaf node size is based on the number of episodes with corresponding diagnoses codes co-occurring with F90. Figure~\ref{fig:graph_preschooler_F_ADHD} illustrates these graphs at the stage of changing the granularity of the diagnosis code from the detailed original level in the ICD-10 hierarchy to the much broader level 1. Figure~\ref{fig:combined_level3} shows the co-occurrence of disorders with F90 at level 3 across all six patient groups, highlighting differences in size and diversity. Figure~\ref{fig:graph_preschooler_F_ADHD} demonstrates that the level 3 graphs are adequately informative and easy to interpret visually. Each diagnosis code is clearly visible in the graph. In contrast, the original level appears messier, and level 1 is too broad.
Figure \ref{fig:combined_level3} shows 
most co-occurring disorders with F90. In males, learning difficulties (F81) are prevalent in teenagers and in middle childhood, and Tic disorders (F95) and disorders of social functioning (F94) are seen in preschoolers. In females, F81 is prevalent in teenagers and F95 in middle childhood and preschoolers. 

All the patient trajectory pattern plots and F90 patient co-occurrence graphs are in \hyperref[app:A]{Supplementary Material}.

\subsection{Clinical insights and evaluation}
We gathered feedback on the understandability and utility of our visualizations with five CAMHS clinicians. Three of these clinicians were involved in the data preprocessing and visualization design, while the remaining two saw the plots and graphs for the first time. Their feedback, collected through interviews and questionnaires (See Supplementary Material\ref{app:A}), focused on the clarity and usefulness of the visualizations. Understandability was measured based on the clarity of the insight and description obtained. Initially, clinicians observed no significant care pattern differences between `ADHD' and `no ADHD' patients, but they did note a higher presence of cluster 2 (purple) in ADHD patients and cluster 1 (green) in `no ADHD' patients. They also understood most of the visualization elements we used (like sizes, labels, and colors).  However, their understanding improved over time with familiarity and discussion.

Preschoolers showed fewer CAMHS episodes, suggesting care supplementation by external services (primary care or educational support), whereas older groups demonstrated intense and consistent CAMHS engagement. Tourette's syndrome (F952) was co-occurring predominantly with F90 in all preschool patients, as well as in middle childhood and teenage male groups. Preschoolers also show autism spectrum disorder (F849) and reactive attachment disorder (F941), highlighting the importance of early multidisciplinary intervention. Learning difficulties (F810, F813) and social challenges (F845) were prevalent in middle childhood, indicating evolving patient needs.  Regarding utility, clinicians reflected that these visualizations can be useful, although they are not yet sure of their specific purpose in some cases. 

However, the functionality we developed for illustrating the information was clear and easily understood by the clinicians, enabling them to relate observations within and across the plots and graphs. For example, they were able to identify that teenage females exhibited more disorders related to depression (F321), self-harm (X60-X84 \cite{finnekode_norge}), and school difficulties (F810),  which aligns with the sustained CAMHS support shown by Figure \ref{fig:patternsplot}. The visualizations aroused their curiosity in different areas, suggesting potential research questions that visualizations could help answer. However, they also highlighted some shortcomings in the visualizations: 1) the need for strong captions, introducing all the elements in the plot or graph that need a description (like what `nan' means). 2) Clinical/technical background knowledge was sometimes needed to understand the visualization. They struggled with understanding concepts like clusters, recognizing the medication codes, or the meaning of letters in diagnostic codes. 3) Readability decreases whenever lots of data is visualized in a plot or graph. The co-occurrence plots in full detail (original level) were too crowded to be understandable or useful, and the least detailed level (level 1) gave rise to perceived data loss and confusion. The ICD10 level 3 seemed to be the most useful. 4) The most significant data points should be emphasized and prioritized in data visualization rather than presenting all of the data.

\section{Conclusion}
We made and evaluated patient trajectory, trajectory pattern plots, and F90 co-occurrence graphs based on the two main criteria: understandability and utility. They generally appeared to be effective in providing simplicity, demonstrating patient demographics and episode of care insight in a user-friendly manner, thus facilitating the decision-making process in CAMHS. Our study demonstrated that visualizing patient timelines as trajectories with the most significant information and using graphs for co-occurring disorders aids clinicians in understanding patient histories and predicting outcomes. These visualizations clarify the results of data analytics and machine learning, particularly the intricacies of clustering algorithms, making the data model results more understandable and useful to clinicians. They also set the stage for continuous improvement of data visualization techniques in future research.
Despite these findings, the visualizations were tested with a relatively small group of clinicians, and the data's scope was limited to a single geographic region. Future research should explore applying these visualization techniques to more diverse and comprehensive datasets from various regions. Additionally, iterative refinement is needed, involving usability and user experience evaluations with a larger group of clinicians and exploring other methods to enhance the understandability and utility of the graph visualizations.
\section*{Acknowledgment}
We acknowledge the Norwegian Research Council for funding the IDDEAS project (grant no. 269117) and the Liaison Committee between the Central Norway Regional Health Authority (RHA) and the Norwegian University of Science and Technology (NTNU) (grant number 30233).  
\appendices

\section*{Supplementary Material\label{app:A}}

Available at: 
\url{https://doi.org/10.5281/zenodo.14055549}

\bibliographystyle{IEEEtran}
\bibliography{sample-base}

\end{document}